\documentclass[letter]{aa}  

\usepackage{graphicx}
\usepackage{txfonts}
\usepackage[pdftex]{hyperref}
\hypersetup{colorlinks=true, citecolor=blue, linkcolor=blue}
\begin{document} 

   \title{A neural network approach to determining photometric metallicities of M-type dwarf stars}
   \titlerunning{}
   \authorrunning{C.~Duque-Arribas et al.}

   \author{C.~Duque-Arribas\inst{1},
           H.~M.~Tabernero\inst{2,3},
           D.~Montes\inst{1},
           J.~A.~Caballero\inst{4}
           \and
           E.~Galceran\inst{1}
          }

   \institute{Departamento de F{\'i}sica de la Tierra y Astrof{\'i}sica \& 
           IPARCOS-UCM (Instituto de F\'{i}sica de Part\'{i}culas y del Cosmos de la UCM), 
           Facultad de Ciencias F{\'i}sicas, Universidad Complutense de Madrid, 28040 Madrid, Spain\\
           \email{chrduque@ucm.es}
           \and
           Institut de Ciències de l’Espai (ICE, CSIC), Campus UAB, c/ de Can Magrans s/n, 08193 Cerdanyola del Vallès, Barcelona, Spain
           \and
           Institut d'Estudis Espacials de Catalunya (IEEC), C/Esteve Terradas, 1, Edifici RDIT, Campus PMT-UPC, 08860 Castelldefels (Barcelona), Spain
           \and
           Centro de Astrobiolog\'ia (CSIC-INTA), ESAC Campus, Camino bajo del castillo s/n, 28692 Villanueva de la Ca\~nada, Madrid, Spain}

   \date{Received 24 March 2025 / Accepted 18 May 2025}

 
  \abstract
   {M dwarfs are the most abundant stars in the Galaxy and serve as key targets for stellar and exoplanetary studies. It is particularly challenging to determine their metallicities because their spectra are complex. For this reason, several authors have focused on photometric estimates of the M-dwarf metallicity. Although artificial neural networks have been used in the framework of modern astrophysics, their application to a photometric metallicity estimate for M dwarfs remains unexplored.}
   {We develop an accurate method for estimating the photometric metallicities of M dwarfs using artificial neural networks to address the limitations of traditional empirical approaches.}
   {We trained a neural network on a dataset of M dwarfs with spectroscopically derived metallicities. We used eight absolute magnitudes in the visible and infrared from \textit{Gaia}, 2MASS, and WISE as input features. Batch normalization and dropout regularization stabilized the training and prevented overfitting. We applied the Monte Carlo dropout technique to obtain more robust predictions.}
   {The neural network demonstrated a strong performance in estimating photometric metallicities for M dwarfs in the range of $-0.45\leq\text{[Fe/H]}\leq +0.45$\,dex and for spectral types as late as M5.0\,V. On the test sample, the predictions showed uncertainties down to $0.08$\,dex. This surpasses the accuracy of previous methods. We further validated our results using an additional sample of 46 M dwarfs in wide binary systems with FGK-type primary stars with well-defined metallicities and achieved an excellent predictive performance that surpassed the $0.1$\,dex error threshold.}
   {This study introduces a machine-learning-based framework for estimating the photometric metallicities of M dwarfs and provides a scalable data-driven solution for analyzing large photometric surveys. The results outline the potential of artificial neural networks to enhance the determination of stellar parameters, and they offer promising prospects for future applications.}


   \maketitle
%

\defcitealias{Birky2020ApJ...892...31B}{B20}
\defcitealias{Bonfils2005A&A...442..635B}{B05}
\defcitealias{Johnson2009ApJ...699..933J}{JA09}
\defcitealias{Neves2012A&A...538A..25N}{N12}
\defcitealias{Mann2013AJ....145...52M}{M13}
\defcitealias{Davenport2019RNAAS...3...54D}{DD19}
\defcitealias{Rains2021MNRAS.504.5788R}{R21}

\section{Introduction} 

Knowledge of stellar parameters and in particular, the chemical composition of stars in the Milky Way, is crucial for several areas of astrophysics. M dwarfs are the most common type of star in our Galaxy by far. They constitute over $75\,\%$ of all stars \citep{Henry2006AJ....132.2360H,Winters2015AJ....149....5W,Reyle2021A&A...650A.201R}. Studies of the metallicity of these cool stars have been made regarding the chemical and dynamical evolution of our Galaxy \citep{Bahcall1980ApJS...44...73B,Reid1997PASP..109..559R,Chabrier2003PASP..115..763C,Ferguson2017ApJ...843..141F} and exoplanet discoveries and characterization \citep[e.g.][]{Nutzman2008PASP..120..317N,Shields2016PhR...663....1S,Reiners2018A&A...609L...5R,Trifonov2018A&A...609A.117T,Luque2019A&A...628A..39L,Caballero2022A&A...665A.120C}.

Nevertheless, determining the metallicity of M dwarfs still remains a challenging task because their spectra are complex. They are dominated by strong molecular features that erode the stellar continuum \citep{Allard1997ARA&A..35..137A,VanEck2017A&A...601A..10V,Passegger2018A&A...615A...6P,Marfil2021A&A...656A.162M}. For this reason, several studies relied on photometric data to estimate the metallicity of these stars, using techniques such as frequentist or Bayesian statistics, $k$-nearest neighbors, or Gaussian-process regressors to provide photometric calibrations for the M-dwarf metallicities \cite[e.g.][]{Bonfils2005A&A...442..635B,Johnson2009ApJ...699..933J, Neves2012A&A...538A..25N, Mann2013AJ....145...52M, Davenport2019RNAAS...3...54D, Rains2021MNRAS.504.5788R, Duque-Arribas2023ApJ...944..106D}. 

Our study proposes a novel method for estimating photometric metallicities of M dwarfs using artificial neural networks (ANNs). The literature knows several ANNs that were applied to astrophysical studies, including the determination of stellar parameters from the Sloan Digital Sky Survey--III APOGEE\footnote{Apache Point Observatory Galactic Evolution Experiment} spectra \citep{Fabbro2018MNRAS.475.2978F} or from CARMENES\footnote{Calar Alto high-Resolution search for M dwarfs with Exoearths with Near-infrared and optical Echelle Spectrographs; \url{https://carmenes.caha.es/index.html}} spectra \citep{Passegger2020A&A...642A..22P, BelloGarcia2023A&A...673A.105B, MasBuitrago2024A&A...687A.205M},  evolutionary states of red giants from asteroseismology \citep{Hon2017MNRAS.469.4578H}, star-galaxy classification \citep{Kim2017MNRAS.464.4463K}, and a stellar spectral classification \citep{Kheirdastan2016Ap&SS.361..304K}. Relatively few studies have explored the use of ANNs for deriving photometric metallicities, however. \cite{Whitten2019A&A...622A.182W} estimated effective temperatures and metallicities for stars hotter than $T_\text{eff}>4500$\,K using J-PLUS\footnote{Javalambre-Photometric Local Universe Survey} photometry \citep{Cenarro2019A&A...622A.176C}. \cite{Fallows2022MNRAS.516.5521F} estimated metallicities of red giant stars from \textit{Gaia} EDR3 \citep{Gaia2016A&A...595A...1G, Gaia2021A&A...649A...1G} and the 2MASS\footnote{Two Micron All-Sky Survey} \citep{Skrutskie2006AJ....131.1163S} and WISE\footnote{Wide-field Infrared Survey Explorer} \citep{Wright2010AJ....140.1868W} surveys, and \cite{Molina-Jorquera2024A&A...691A.144M} relied on S-PLUS\footnote{Southern Photometric Local Universe Survey} photometry \citep{MendesdeOliveira2019MNRAS.489..241M} to classify giant and dwarf stars and to determine the metallicity of red giants. Finally, \cite{FerreiraLopes2025A&A...693A.306F} estimated the stellar parameters for about five million stars from S-PLUS multiband photometry, although the authors indicated that the results for stars with $T_\text{eff}<4000$\,K should be interpreted with caution. Despite these advancements, no study to date has specifically focused on applying ANNs to estimate photometric metallicities for M dwarfs as a standalone population. We present an ANN that we used to estimate the metallicity of M-type dwarf stars based on their visible and infrared photometry.


\section{Method}

\subsection{Data and preprocessing}

\begin{table}
    \centering
    \small
    \caption{Data-filtering criteria applied to astrophotometric data.}
    \begin{tabular}{lc} 
    \hline
    \hline
    \noalign{\smallskip}
    Survey & Filter \\
    \noalign{\smallskip}
    \hline
    \noalign{\smallskip}
    \textit{Gaia} EDR3 & $\texttt{parallax\_over\_error > 10}$ \\
     & $\texttt{ruwe < 1.4}$ \\
     & $\texttt{photo\_g\_mean\_flux\_over\_error > 50}$ \\
     & $\texttt{photo\_bp\_mean\_flux\_over\_error > 20}$ \\
     & $\texttt{photo\_rp\_mean\_flux\_over\_error > 20}$ \\ 
    \noalign{\smallskip}
    \hline
    \noalign{\smallskip}
    2MASS & $\texttt{Qfl = AAA}$ \\ 
    \hline
    \noalign{\smallskip}
    CatWISE2020 & $\texttt{qph = AA**}$ \\
    \noalign{\smallskip}
    \hline
    \end{tabular}
    \tablefoot{
    {\tt Qfl} and {\tt qph} are the quality flags in the 2MASS $JHK_s$ bands and in the WISE $W1W2W3W4$ bands, respectively.
    }
    \label{tab:data_filtering}
\end{table}

The proposed approach was carried out on the stellar sample described by \cite{Duque-Arribas2023ApJ...944..106D}. They made use of the sample of 5875 early- and intermediate-type M dwarfs presented by \cite{Birky2020ApJ...892...31B}, with stellar parameters derived from APOGEE spectra \citep{Majewski2017AJ....154...94M, Abolfathi2018ApJS..235...42A} using \texttt{The Cannon} \citep{Ness2015ApJ...808...16N}, and they cross-matched these stars with the \textit{Gaia} DR3 \citep{Gaia2023A&A...674A...1G}, 2MASS, and CatWISE2020 \citep{Marocco2021ApJS..253....8M} catalogs. Then, they filtered the resulting sample based on the photometric and astrometric quality criteria compiled in Table~\ref{tab:data_filtering} while removing young and evolved objects. They obtained a final sample of 4919 M dwarfs. 

This sample was then randomly divided into three subsets: $60\%$ for training, $20\%$ for validation, and $20\%$ for testing, that is, 2951, 984 and 984 stars, respectively. The first subset was used to train the ANN, the second subset to estimate its accuracy and tune the hyperparameters, and the third subset to test the ANN. 

Following the approach of \cite{Duque-Arribas2023ApJ...944..106D}, we evaluated the predictive performance of the ANN using a separate dataset of FGK+M binary systems from \cite{Montes2018MNRAS.479.1332M}. The primary stars have well-determined stellar atmospheric parameters derived from high-resolution spectroscopy with the equivalent width method \citep{Tabernero2019A&A...628A.131T}, and it is reasonable to assume that the M-dwarf companions share the same chemical composition \citep{Desidera2006A&A...454..581D, Andrews2018MNRAS.473.5393A}. 

To prepare the dataset, we cross-matched the M-dwarf companions with the \textit{Gaia} DR3, 2MASS, and CatWISE2020 catalogs and identified 115 stars in common. We then applied the same data-filtering criteria as described earlier and constrained stars with parameters similar to the \cite{Birky2020ApJ...892...31B} sample. Specifically, we limited the sample to stars with $-0.45\leq\text{[Fe/H]}\leq 0.45$\,dex, $1.85$\,mag $\leq G-J\leq$ $3.10$\,mag, and $-0.10$\,mag $\leq W1-W2\leq$ $0.24$\,mag, which correspond to early- and intermediate-M dwarfs (spectral types M0.0\,V to M5.0\,V approximately; \citealt{Cifuentes2020A&A...642A.115C}). After applying these constraints, we obtained a final sample of 46 M-dwarf companions for testing the ANN on a completely independent dataset.

\subsection{ANN architecture}

\begin{figure*}
    \centering
	\includegraphics[width=1.55\columnwidth]{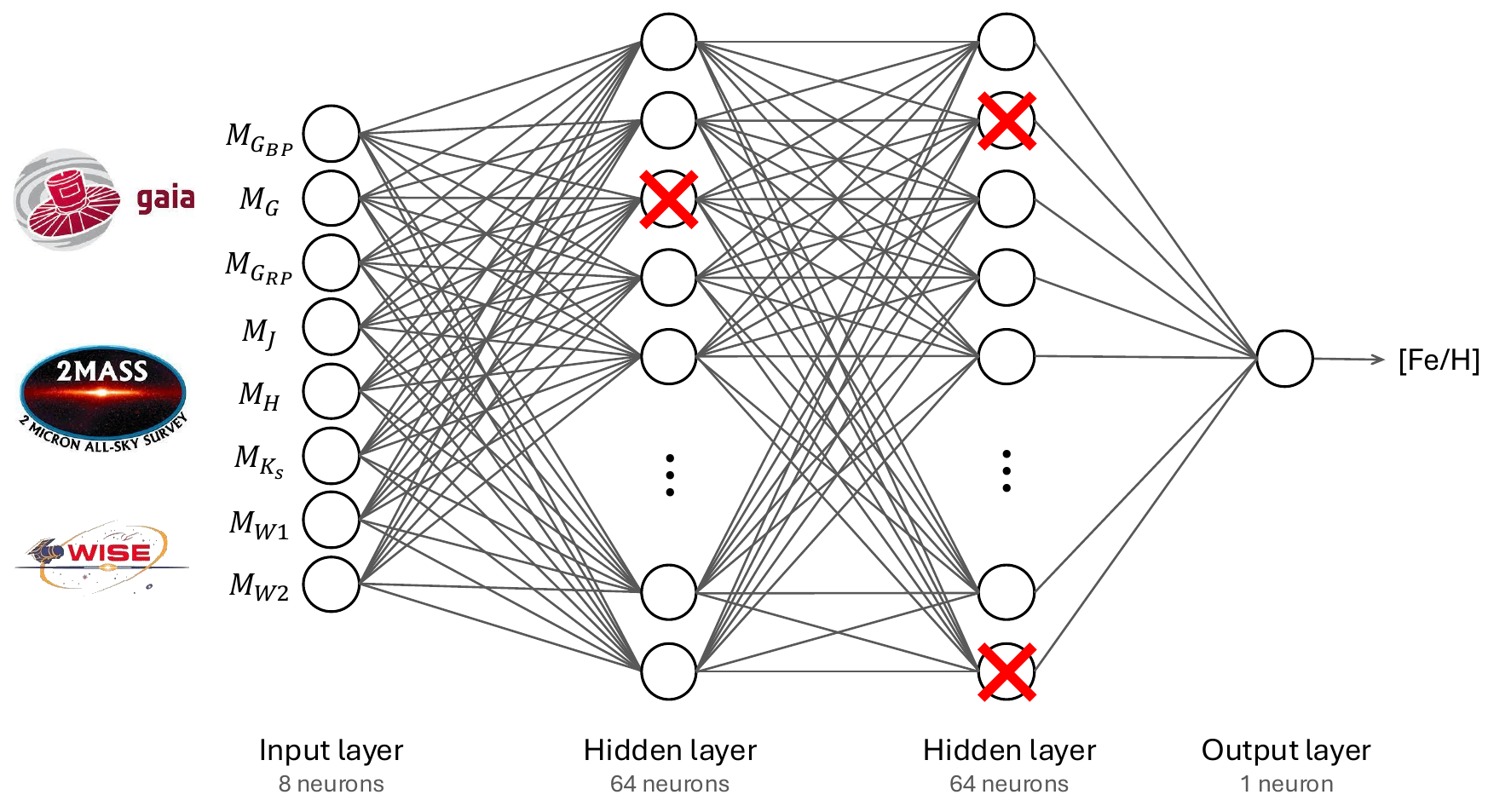}
    \caption{Schematic representation of our ANN architecture.
    The red crosses indicate dropout neurons.}
    \label{fig:nn_architecture}
\end{figure*}

We employed the \texttt{Keras}\footnote{\url{https://keras.io/about/}} API, which operates on the \texttt{TensorFlow}\footnote{\url{https://www.tensorflow.org/}} framework, to build our model. Hyperparameter optimization, in our case, for the number of hidden layers, the number of neurons per layer, and the learning rate, was performed using the \texttt{RandomizedSearchCV} class from the \texttt{scikit-learn}\footnote{\url{https://scikit-learn.org/stable/}} library. This approach efficiently explores the hyperparameter space by sampling it randomly, allowing for the identification of near-optimal configurations in less time compared to exhaustive methods, such as the grid search. The optimization process used a $k$-fold cross-validation, with four folds in our implementation, and it evaluated the model based on a modified mean squared error (MSE) as the loss metric.

The modification of the MSE was included to mitigate the overabundance of solar-metallicity stars in the sample, as noted by \cite{Fallows2022MNRAS.516.5521F}. Otherwise, the ANN might inadvertently interpret this imbalance as a trend in the data, which would lead to biased predictions. To counteract this, we included a weighting factor $W=4 \times |\text{[Fe/H]}| +0.5$ as a linear multiplier on the MSE. This weighting reduced the influence of the stars with $\text{[Fe/H]}\approx 0$ while it linearly increased the effect for objects with higher or lower metallicities. This ensured a more balanced contribution throughout the entire metallicity range. We tested alternative weighting strategies, including quadratic and exponential functions, but found that they did not improve the results. The linear form was chosen for its simplicity and effectiveness. The specific coefficients were determined empirically: The slope ensured a gradual yet sufficient increase in weight as the metallicity deviated from solar, while the intercept of $0.5$ prevented complete suppression of the most abundant metallicities. This balance was found to stabilize training and improve model generalization, in particular, for stars at the extremes of our metallicity range.

The architecture of the neural network was chosen based on the empirical performance during the hyperparameter tuning. We explored a range of configurations that are summarized in Table~\ref{tab:hyperparameters}. Our final architecture, with two hidden layers of 64 neurons each, consistently achieved the lowest validation loss without signs of overfitting. Simpler architectures showed higher validation errors, while more complex architectures offered no additional performance gain and occasionally led to overfitting.

\begin{table}
    \centering
    \small
    \caption{Summary of the tested ANN hyperparameters.}
    \begin{tabular}{lc} 
    \hline
    \hline
    \noalign{\smallskip}
    Hyperparameter & Explored values \\ 
    \noalign{\smallskip}
    \hline
    \noalign{\smallskip}

    Number of hidden layers & 1, 2, 3, 4, 5 \\
    Neurons per layer & 8, 16, 32, 64, 128, 256 \\
    Learning rate & reciprocal(1$\cdot$10$^{-4}$, 1$\cdot$10$^{-2}$) \\
    Dropout rate & $0.1$, $0.2$, $0.3$, $0.4$, $0.5$ \\
    
    \noalign{\smallskip}
    \hline
    \end{tabular} 
    \label{tab:hyperparameters}
\end{table}

The input layer incorporated eight features, each corresponding to an absolute magnitude using the \textit{Gaia} parallaxes: $M_{G_\text{BP}}$,
$M_G$, and  $M_{G_\text{RP}}$ from \textit{Gaia}, $M_J$, $M_H$, and $M_{K_\text{S}}$ from 2MASS, and $M_{W1}$, and $M_{W2}$ from CatWISE2020. The hidden layers, fully connected, contained 64 neurons per layer. The output layer consisted of a single neuron for regression. The network was trained using the Adam optimizer (a stochastic gradient descent method; \citealt{kingma2017adammethodstochasticoptimization}) with a learning rate of $2\cdot10^{-4}$. For the learning rate, we sampled values from a reciprocal (log-uniform) distribution in the range [1$\cdot$10$^{-4}$, 1$\cdot$10$^{-2}$], as this approach gives greater weight to lower values, which are more likely to yield stable convergence during training. A schematic representation of the neural network is displayed in Fig.~\ref{fig:nn_architecture}.

The neurons in the input and hidden layers incorporated linear activation functions, whereas the output layer adopted a hyperbolic tangent (tanh) activation function. This choice constrained the predictions to the range $(-1,+1)$\,dex. Because the training sample spans a narrower metallicity range ($-0.45$\,dex to $+0.45$\,dex), this activation ensured that predictions remained bounded within a safe interval during training. It also implies a limitation, however: The model is not suited for extrapolating beyond the training range. For stars with true metallicities outside this domain, the output will saturate toward $\pm1$, reflecting the asymptotic behaviour of the tanh function. These predictions should not be interpreted as accurate estimates, but rather as an indication that the input is outside the reliable operating range of the model.

We also incorporated batch normalization after the input and hidden layers. This technique normalizes activations during training, stabilizing and accelerating convergence by mitigating internal covariate shifts \citep{Ioffe2015arXiv150203167I}.
Additionally, dropout was implemented after the hidden layers, with the dropout rate included as a tunable hyperparameter. This technique is discussed in greater detail in the following subsection.

\subsection{Monte Carlo dropout}

Dropout is one of the most popular regularization techniques for neural networks that aims at mitigating overfitting during the training stage by randomly dropping a fraction of the neurons \citep{Hinton2012arXiv1207.0580H, Srivastava2014JMLR:v15:srivastava14a}. During every training step, each neuron has a certain probability that is included as a tuneable hyperparameter of being ignored but being active during the next step. By applying this technique, the ANN learns as a cohesive unit and avoids over-reliance on any particular neuron. We applied dropout to the layers of our neural network and set the dropout rate at $20\%$.

In the classical dropout, neurons are no longer dropped after the training stage. \cite{Gal2016pmlr-v48-gal16} introduced a technique called Monte Carlo (MC) dropout, however, which enhances the performance of a trained dropout model and provides a more accurate measure of the model uncertainty. This is achieved by ensuring that the dropout layers remain active after the training, which produces a distribution of predictions instead of a single estimate. Averaging over these multiple predictions with dropout provides a Monte Carlo estimate that is more reliable than a single prediction. \cite{Gal2016pmlr-v48-gal16} also established a deep connection between dropout networks and the approximate Bayesian inference.

\section{Results and discussion}

\begin{figure}
    \centering
    \includegraphics[width=0.99\columnwidth]{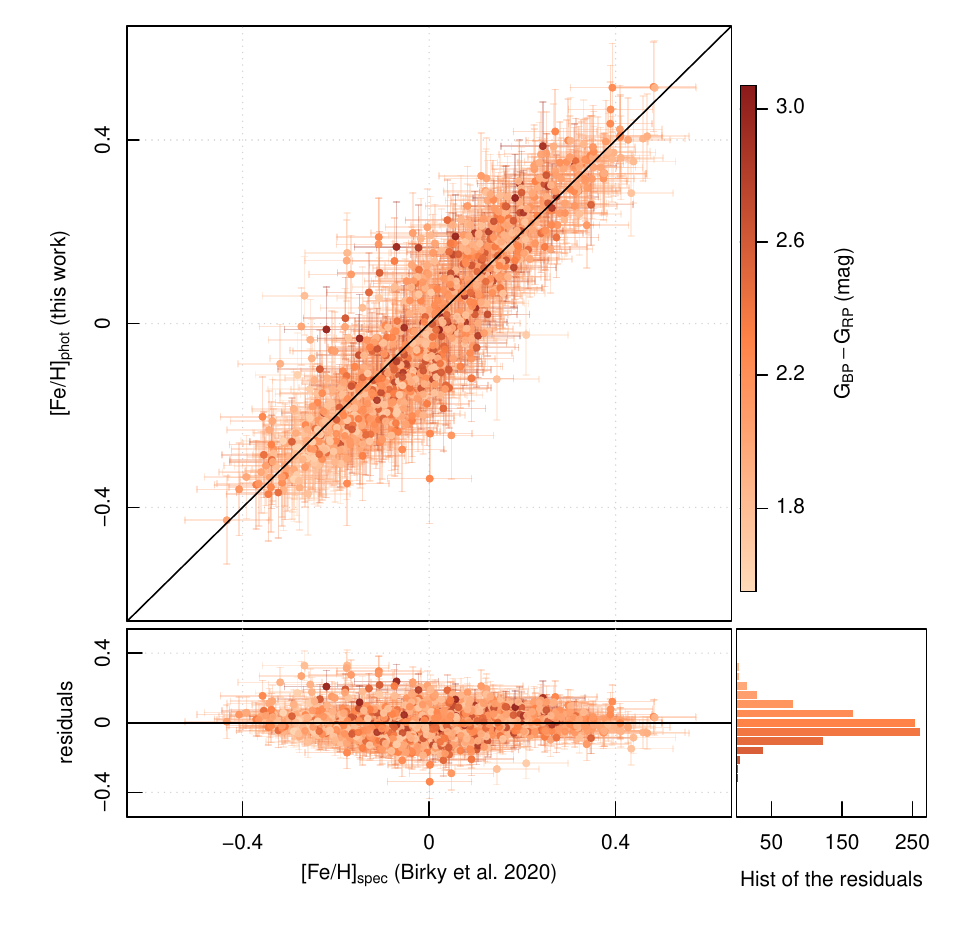}
    \caption{Comparison between spectroscopic metallicities reported by \cite{Birky2020ApJ...892...31B} and photometric estimates by our ANN.
    The 984 stars from the test sample and their corresponding residuals are color-coded by $G_\text{BP}-G_\text{RP}$ color (darker symbols show cooler stars, and lighter symbols show warmer stars). The solid lines denote the 1:1 relation, and the residuals are zero. In the bottom right corner we represent the histogram of the residuals.}
    \label{fig:FeH_predicted}
\end{figure}

\begin{figure}
    \centering
	\includegraphics[width=0.99\columnwidth]{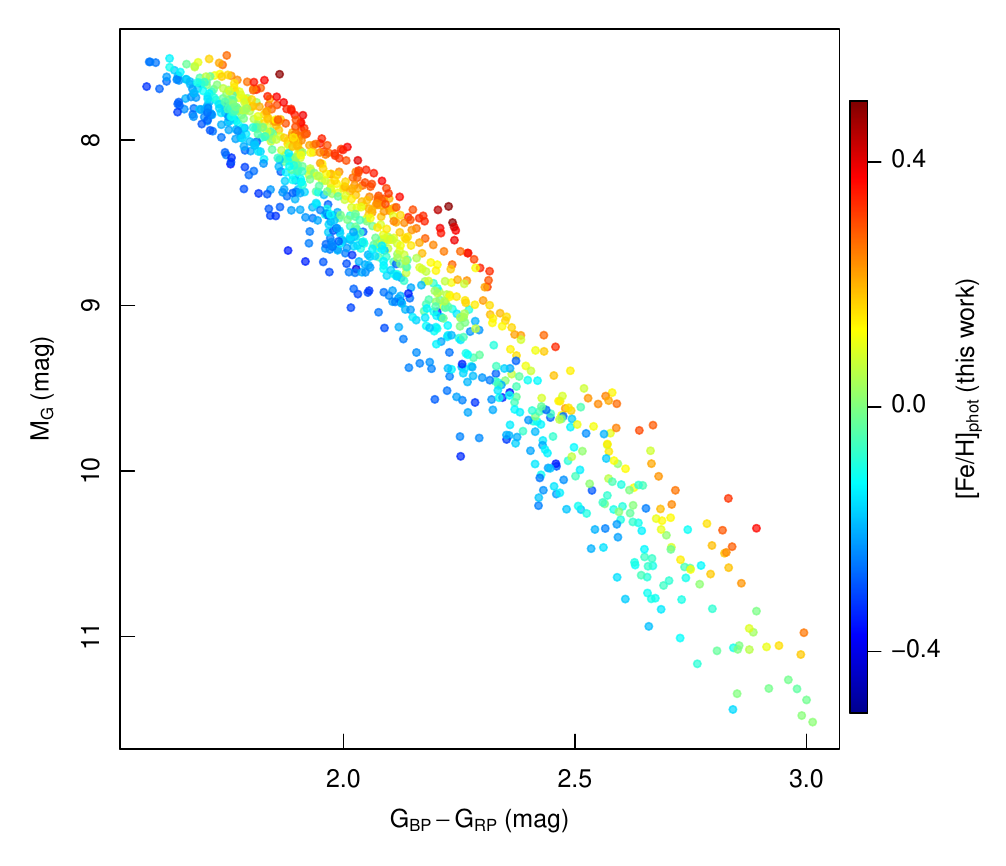}
    \caption{Color-magnitude diagram of the 984 stars from the test sample, color-coded by the metallicity estimates provided by our ANN.}
    \label{fig:cmd}
\end{figure}

After training the ANN and optimizing its hyperparameters using the training and validation datasets, we evaluated its performance on the test sample. As outlined in the previous section, we applied MC dropout to generate 100 stochastic forward passes per star. This provided more robust estimates and an uncertainty measure for these predictions. 
We also generated 500 samples, but the predictions did not change.

In Fig.~\ref{fig:FeH_predicted} we compare the spectroscopic metallicity values reported by \cite{Birky2020ApJ...892...31B} and the photometric metallicity predictions generated by our ANN for the test subsample, color-coded by $G_\text{BP}-G_\text{RP}$, which is a proxy for the stellar effective temperature \citep{Cifuentes2020A&A...642A.115C}. Most of the estimated metallicities closely align with the one-to-one relation. This indicates the reliability of the model. Furthermore, as expected, no significant correlation is observed between the predicted metallicities and effective temperature. The prediction residuals exhibit a standard deviation of $0.08$\,dex, which is compatible with the values reported by \cite{Duque-Arribas2023ApJ...944..106D} using a Bayesian approach.

In Fig.~\ref{fig:cmd} we present the color-magnitude diagram for the test subsample, color-coded by metallicity estimates from our ANN. The diagram shows the clear recovery of the expected metallicity gradient along the main sequence. This alignment underscores the ability of the ANN to capture and reproduce well-known trends in stellar populations. For a detailed discussion of metallicity gradients in the color-color and color-magnitude diagrams, we refer to \cite{Duque-Arribas2023ApJ...944..106D}.

\begin{figure}
    \centering
	\includegraphics[width=0.85\columnwidth]{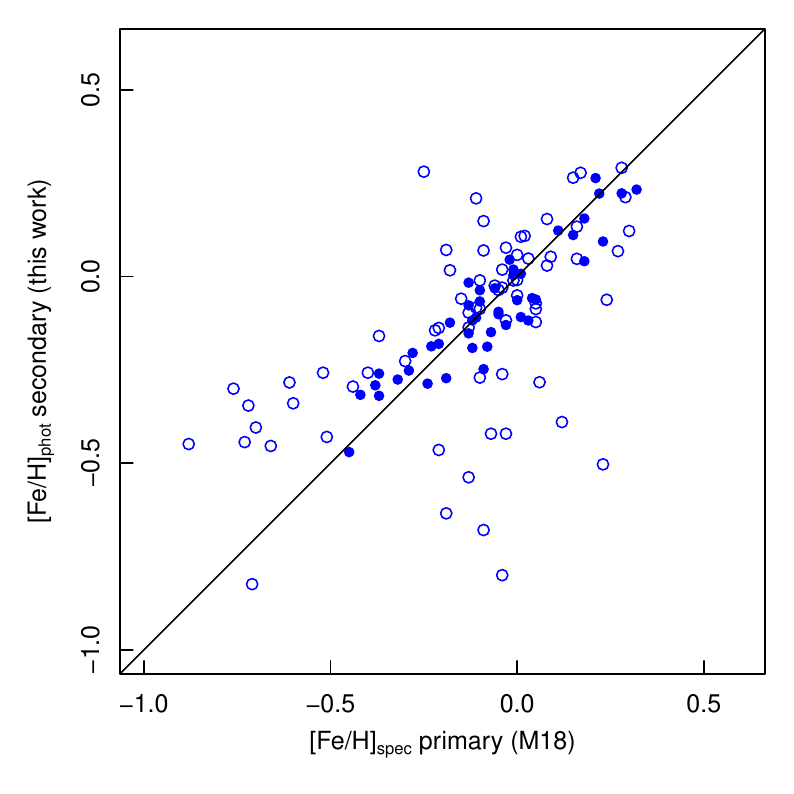}
    \caption{Comparison between spectroscopic metallicity of the FGK-type primary stars measured by \cite{Montes2018MNRAS.479.1332M} and the photometric metallicity estimated for the M-dwarf companions by this work using an ANN. The filled circles represent the 46 good-quality subsample}.
    \label{fig:FeH_comparisons}
\end{figure}

Finally, as an additional test, we applied our ANN to the M-dwarf secondaries in wide binary systems reported by \cite{Montes2018MNRAS.479.1332M}. In Fig.~\ref{fig:FeH_comparisons} we compare the spectroscopic metallicities of the primary stars reported by \cite{Montes2018MNRAS.479.1332M} and the photometrically estimated values for the M-dwarf companions using our ANN. The statistics of the residuals between the spectroscopic and photometric metallicities for the 46 stars with good-quality data, obtained with previous photometric studies found in the literature and with our ANN, are reported in Table~\ref{tab:hist_statistics}. We achieved a strong correlation between photometric and spectroscopic metallicities of the M dwarfs compared with these previous results, as illustrated by Fig.~\ref{fig:comparison_phot_lit}.

Among the most recent studies, both \citet{Rains2021MNRAS.504.5788R} and \citet{Duque-Arribas2023ApJ...944..106D} employed polynomial fits to the main sequence to estimate the photometric metallicities of M dwarfs. The model by \citet{Rains2021MNRAS.504.5788R} relied on \textit{Gaia} and 2MASS photometry alone and omitted the W1$-$W2 color index, which is a known metallicity-sensitive feature \citep{Schmidt2016MNRAS.460.2611S}. This omission likely contributes to its slightly larger dispersion compared to the model by \citet{Duque-Arribas2023ApJ...944..106D} and the present work. Our results are similar to those of \citet{Duque-Arribas2023ApJ...944..106D}, as expected because both studies used the same underlying dataset. The modest improvement achieved here can probably be attributed to the greater flexibility of the ANN relative to the polynomial fitting approach.

\begin{table}
    \centering
    \small
    \caption{Statistics$^a$ of the differences between the  primary [Fe/H]$_\text{spec}$ and the secondary [Fe/H]$_\text{phot}$.}
    \begin{tabular}{l cccc cc} 
    \hline
    \hline
    \noalign{\smallskip}
     & $\overline{x}$ & $\tilde{x}$ & $\sigma_x$ & MAD($x$) & $r$ & $r_{\rm S}$ \\ 
    \noalign{\smallskip}
    \hline
    \noalign{\smallskip}

    B05 & $+0.144$ & $+0.111$ & $0.147$ & $0.105$ & $0.71$ & $0.67$ \\
    JA09 & $-0.064$ & $-0.086$ & $0.206$ & $0.144$ & $0.69$ & $0.66$ \\
    N12 & $+0.071$ & $+0.039$ & $0.149$ & $0.115$ & $0.69$ & $0.66$ \\
    M13 & $+0.043$ & $+0.063$ & $0.144$ & $0.156$ & $0.65$ & $0.64$ \\
    DD19 & $+0.047$ & $+0.056$ & $0.176$ & $0.167$ & $0.53$ & $0.52$ \\
    R21 & $-0.007$ & $+0.001$ & $0.105$ & $0.089$ & $0.84$ & $0.78$ \\
    DA23 & $-0.012$ & $-0.004$ & $0.087$ & $0.081$ & $0.90$ & $0.83$ \\
    This work & $+0.015$ & $+0.001$ & $0.082$ & $0.075$ & $0.92$ & $0.88$ \\
    
    \noalign{\smallskip}
    \hline
    \end{tabular}
    \tablefoot{
    \tablefoottext{a}{Mean ($\overline{x}$), median ($\tilde{x}$), standard deviation ($\sigma_x$), median absolute deviation (MAD($x$)), and Pearson 's ($r$) and Spearman's ($r_{\rm S}$) correlation coefficients, where $x =$ [Fe/H]$_\text{spec}$ -- [Fe/H]$_\text{phot}$ in dex.} 
    }
    \label{tab:hist_statistics}
\end{table}

\section{Conclusions}

We presented a machine-learning framework for estimating the photometric metallicities of M dwarfs, for which we exploited the capabilities of ANNs combined with the accuracy and homogeneity of the visible and infrared photometry from the \textit{Gaia} DR3, 2MASS, and CatWISE surveys. By employing robust techniques such as weighting adjustments in the loss function to mitigate the effects of sample imbalances, MC dropout for the uncertainty estimation, and hyperparameter optimization, we demonstrated that our ANN with two hidden layers can achieve highly accurate and reliable predictions.

We trained the model with the sample presented by \cite{Birky2020ApJ...892...31B}. Its performance on the test sample showed a mean residual of $0.00\pm0.08$\,dex. Additionally, the results confirmed that the predicted metallicities closely match spectroscopic values and exhibit no correlation with the effective temperature. This validated the ability of the model to generalize without introducing biases. The validation on an independent dataset of 46 M dwarfs in FGK+M binary systems from \cite{Montes2018MNRAS.479.1332M} further reinforced the predictive accuracy of the ANN and underscored its versatility when applied to new data. These results agree well with those presented by \cite{Duque-Arribas2023ApJ...944..106D} using a Bayesian linear regression approach.

This ANN-based approach provides a scalable and efficient alternative for processing large photometric datasets. Its ability to infer stellar metallicities using only broadband photometry significantly reduces the reliance on time-intensive spectroscopic observations. This framework can be extended to incorporate additional features or input data to enhance the accuracy and applicability of the predictions. Furthermore, the model could be adapted to other stellar types or parameter determinations, which would broaden its applicability in stellar astrophysics. The Python code is publicly available at GitHub\footnote{\url{https://github.com/chrduque/metamorphosis-NN.git}}.

In summary, this work highlights the potential of machine learning, particularly ANNs, in advancing our ability to determine the stellar metallicity efficiently and accurately. It paves the way for transformative applications in the era of large-scale astronomical surveys.

\begin{acknowledgements}
    We thank the anonymous referee for the instructive and detailed report, which clearly improved our manuscript.
    We acknowledge financial support from the Universidad Complutense de Madrid and the Agencia Estatal de Investigaci\'on (AEI/10.13039/501100011033) of the Ministerio de Ciencia e Innovaci\'on and the ERDF ``A way of making Europe'' through projects 
    PID2019-107427GB-C31, 
    PID2022-137241NB-C4[2,4],	
    and PID2022-138855NB-C31, 
    and the European Research Council (ERC) under the European Union's Horizon Europe programme (ERC Advanced Grant SPOTLESS; no. 101140786).
\end{acknowledgements}

%
\bibliographystyle{aa} 
\bibliography{mybib} 
%

\appendix

\onecolumn
\section{Additional figure}

\begin{figure*}[h!]
    \centering
       \begin{tabular}{cccc}
         \includegraphics[width=0.225\textwidth]{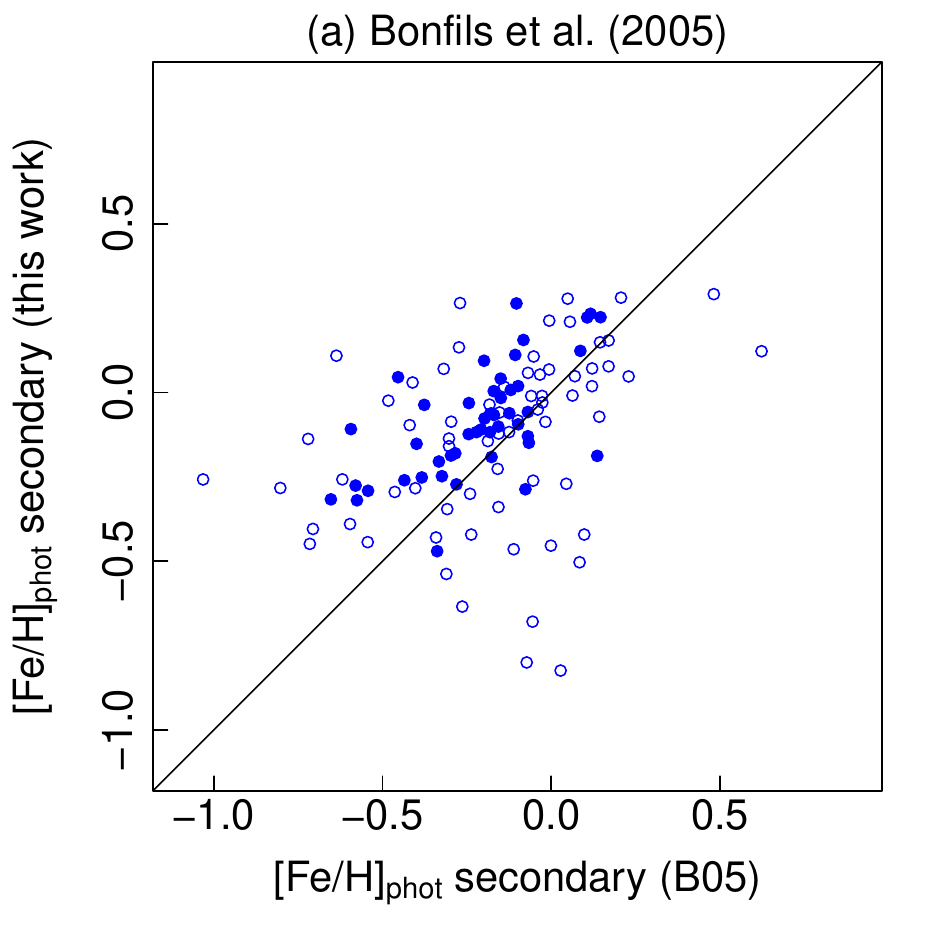} & 
         \includegraphics[width=0.225\textwidth]{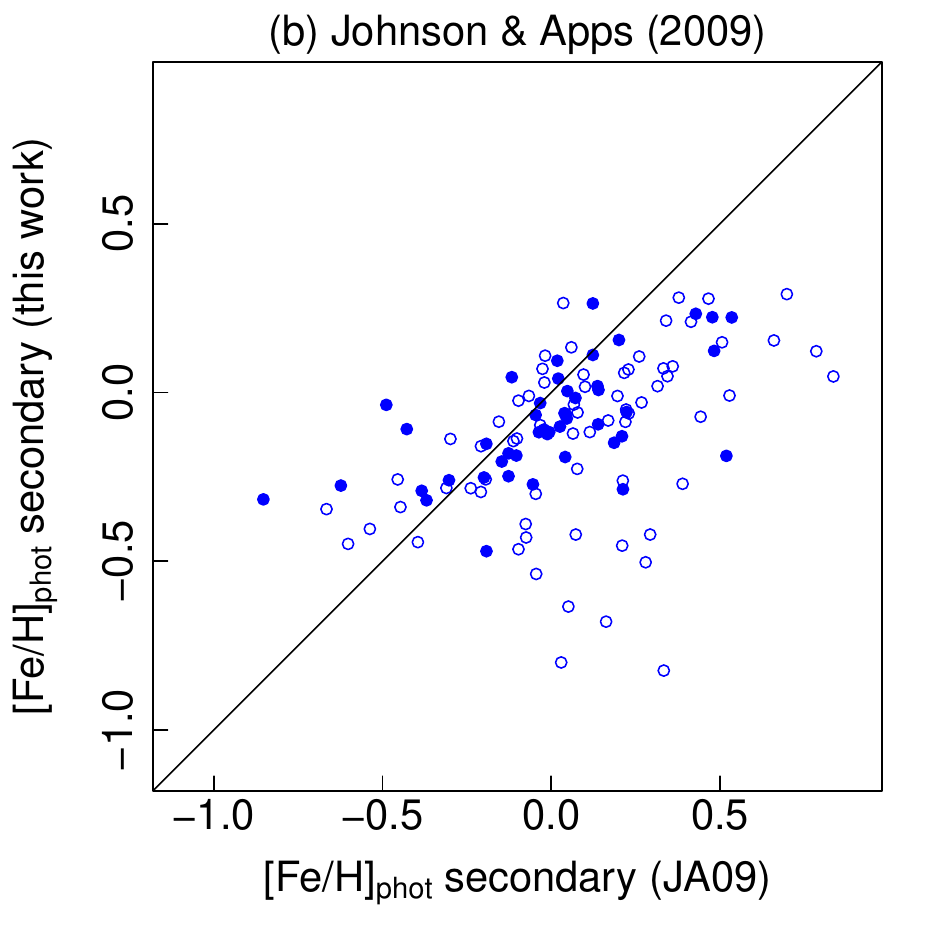} &
         \includegraphics[width=0.225\textwidth]{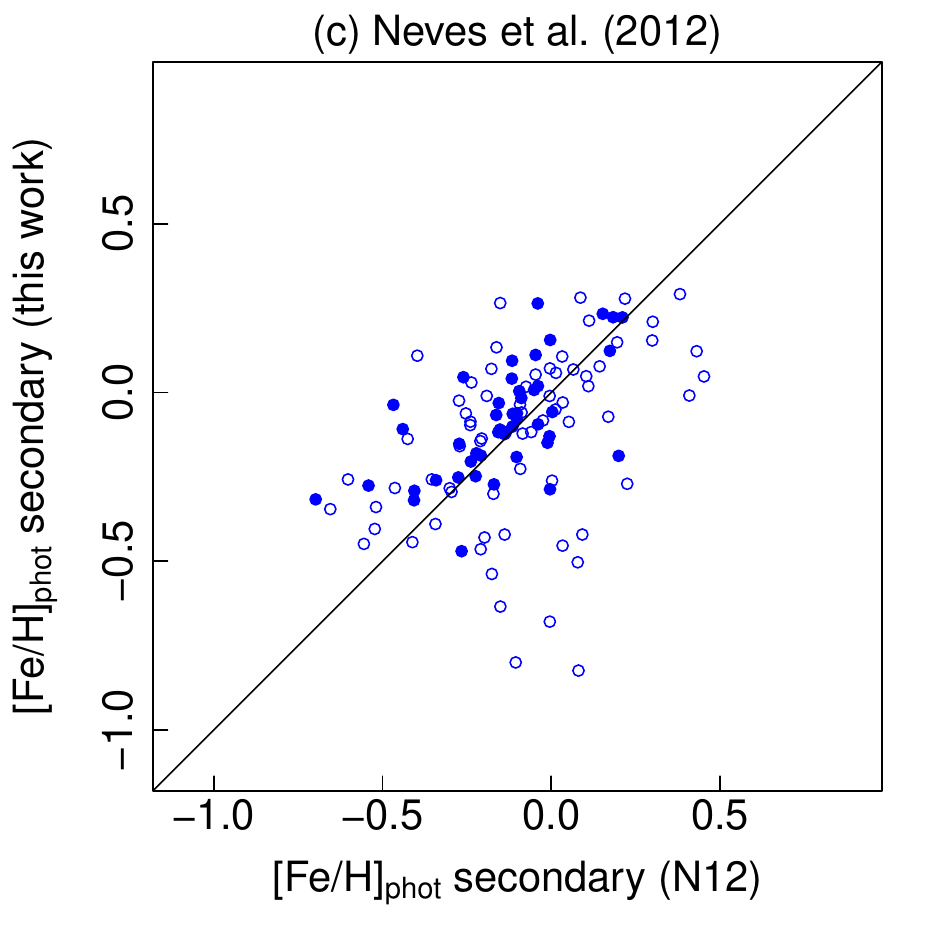} &
         \includegraphics[width=0.225\textwidth]{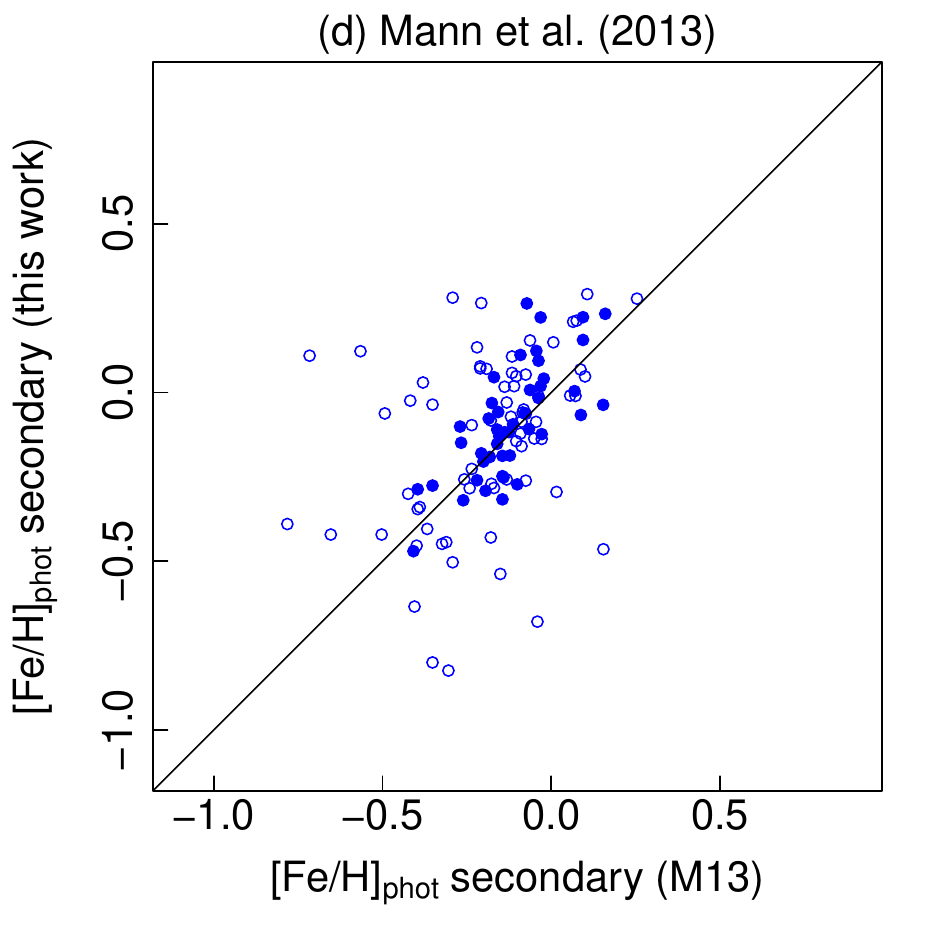}
       \end{tabular}\vspace{2mm}
       
       \begin{tabular}{ccc}
         \includegraphics[width=0.225\textwidth]{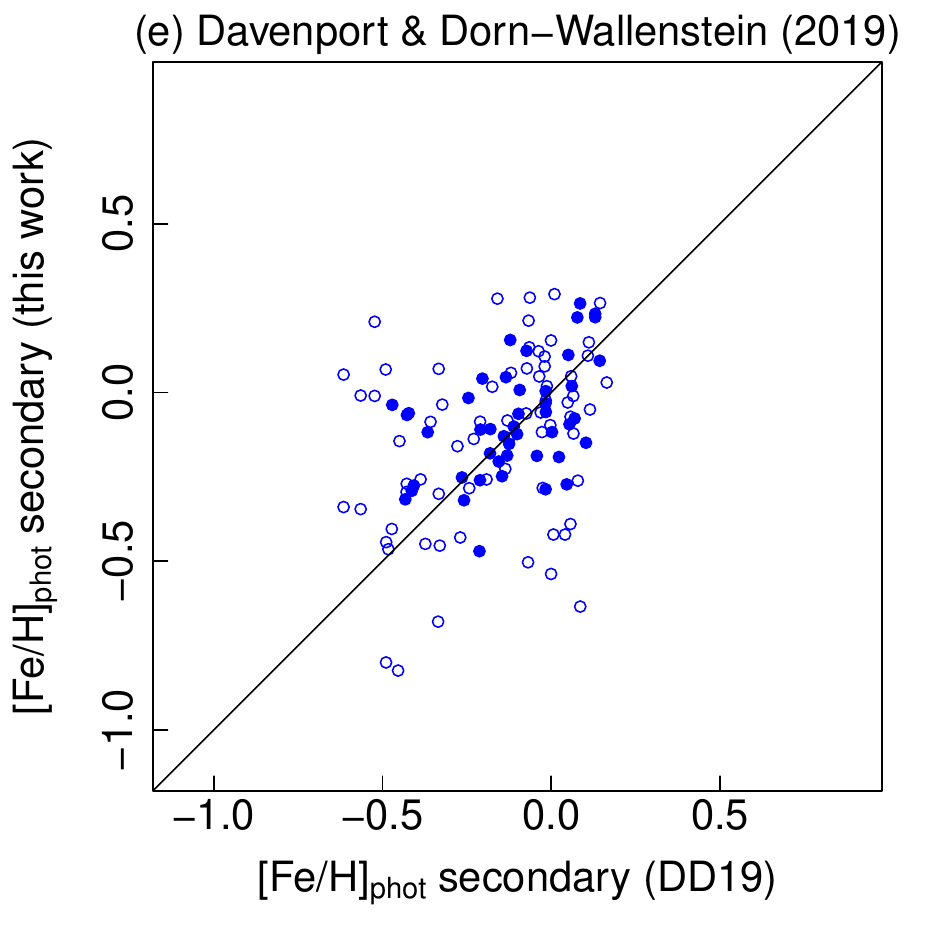} &
         \includegraphics[width=0.225\textwidth]{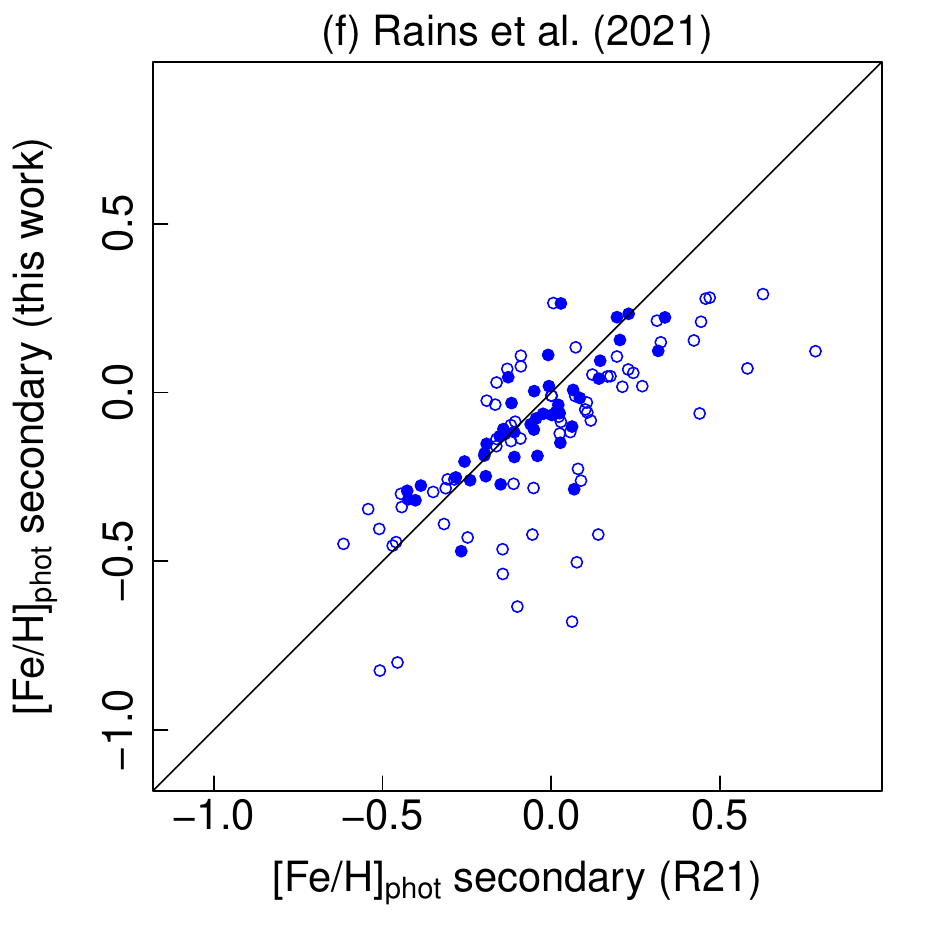} &
         \includegraphics[width=0.225\textwidth]{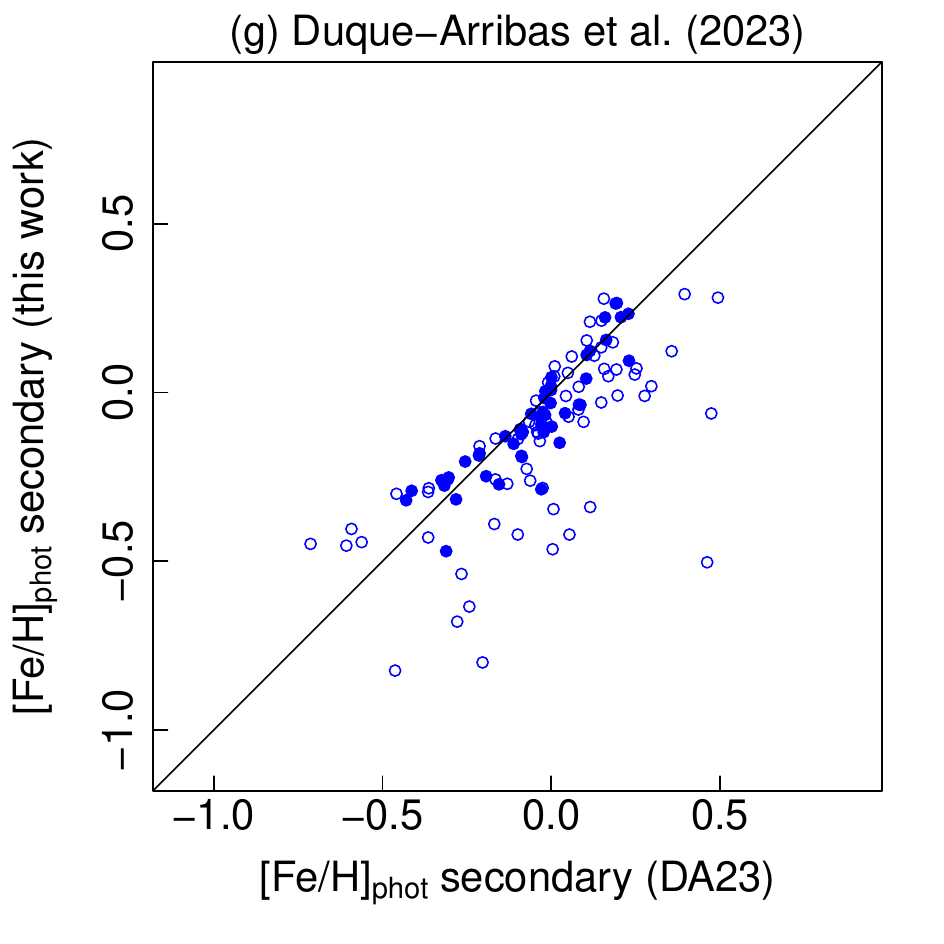}
       \end{tabular}
    \caption{Same as Fig.~\ref{fig:FeH_comparisons}, but for the photometric metallicities estimated by \cite{Bonfils2005A&A...442..635B}, \cite{Johnson2009ApJ...699..933J}, \cite{Neves2012A&A...538A..25N}, \cite{Mann2013AJ....145...52M}, \cite{Davenport2019RNAAS...3...54D}, \cite{Rains2021MNRAS.504.5788R}, and \cite{Duque-Arribas2023ApJ...944..106D}.}   
    \label{fig:comparison_phot_lit}
\end{figure*}

\twocolumn

\end{document}